%% file: main.tex
\documentclass[acmsmall,screen,nonacm]{acmart}

\AtBeginDocument{%
  \providecommand\BibTeX{{%
    Bib\TeX}}}

\setcopyright{none}
\usepackage{enumitem}
\usepackage{amsmath}
\usepackage{algorithmic}
\usepackage[ruled,vlined,linesnumbered]{algorithm2e}
\SetKwInput{KwIn}{Input}
\SetKwInput{KwOut}{Output}
\SetKwRepeat{Repeat}{repeat}{until}
\SetAlFnt{\footnotesize}
\SetAlCapFnt{\footnotesize}
\SetAlCapNameFnt{\footnotesize}
\SetAlgoNlRelativeSize{-1}
\SetAlgoInsideSkip{smallskip}
\DontPrintSemicolon
\usepackage{multirow}
\usepackage[most]{tcolorbox}
\usepackage{array}
\def\BibTeX{{\rm B\kern-.05em{\sc i\kern-.025em b}\kern-.08em
    T\kern-.1667em\lower.7ex\hbox{E}\kern-.125emX}}

\newtcolorbox{summarybox}{
    enhanced,
    colback=gray!15,
    colframe=gray!35,
    boxrule=1.2pt,
    left=1em,
    right=1em,
    top=0.5em,
    bottom=0.5em,
    borderline west={3pt}{0pt}{darkgray},
}

\begin{document}

\input{chapter/000-MetaInfo.tex}

\input{chapter/001-Abstract.tex}

\maketitle

\input{chapter/01-Introduction}
\input{chapter/02-BackgroundMotivation}
\input{chapter/03-Approach}
\input{chapter/05-ExperimentalSettings}
\input{chapter/06-ExperimentalResults}
\input{chapter/07-Discussion}
\input{chapter/08-RelatedWork}
\input{chapter/09-Conclusion}

\bibliographystyle{ACM-Reference-Format}
\bibliography{reference}

\end{document}

%% file: chapter/000-MetaInfo.tex
\title[FeatLens]{FeatLens: Feature-Guided Dependency Retrieval for Repository-Level Code Generation}

\author{Xutian Li}
\affiliation{%
  \department{School of Computer Science}
  \institution{Peking University}
  \city{Beijing}
  \country{China}
}
\email{xtli25@stu.pku.edu.cn}

\author{Bo Xiong}
\affiliation{%
  \department{School of Computer Science}
  \institution{Peking University}
  \city{Beijing}
  \country{China}
}
\author{Yifeng Zhu}
\affiliation{%
  \department{School of Computer Science}
  \institution{Peking University}
  \city{Beijing}
  \country{China}
}
\author{Runbang Yan}
\affiliation{%
  \department{School of Computer Science}
  \institution{Peking University}
  \city{Beijing}
  \country{China}
}
\author{Yanzhen Zou}
\affiliation{%
  \department{School of Computer Science}
  \institution{Peking University}
  \city{Beijing}
  \country{China}
}
\author{Lu Zhang}
\affiliation{%
  \department{School of Computer Science}
  \institution{Peking University}
  \city{Beijing}
  \country{China}
}
\author{Bing Xie}
\affiliation{%
  \department{School of Computer Science}
  \institution{Peking University}
  \city{Beijing}
  \country{China}
}

\renewcommand{\shortauthors}{Li et al.}
\authorsaddresses{Contact: Xutian Li, \href{mailto:xtli25@stu.pku.edu.cn}{xtli25@stu.pku.edu.cn}}

%% file: chapter/001-Abstract.tex
\begin{abstract}
Retrieving task-relevant code dependencies is critical for repository-level code generation. 
Existing approaches obtain dependency context through similarity-based retrieval, repository graphs, or LLM-driven exploration. A central challenge is to identify the dispersed dependencies needed to implement a functional requirement, rather than merely retrieve broadly related code.  
This paper presents FeatLens, a feature-guided dependency retrieval approach for repository-level code generation. FeatLens first builds an offline index linking natural-language feature descriptions to function-level code entities. Given a task requirement and target function location, it selects relevant features and constructs dependency seeds. It then expands the seed graph and applies semantic and structural heuristics to select dependencies for inclusion in the generation context. 
Experiments on DevEval and EvoCodeBench show that FeatLens achieves the highest dependency recall among the evaluated baselines at each fixed retrieval budget, reaching DR@15 values of 0.527 and 0.529, respectively. On DevEval, the retrieved context supports the highest non-oracle Pass@1 with both DeepSeek-V3.2 and GPT-5-mini. These results are obtained without LLM calls in the online dependency retrieval stage.
\end{abstract}

\ccsdesc[500]{Software and its engineering~Software functional properties}
\ccsdesc[500]{Software and its engineering~Software maintenance tools}
\ccsdesc[300]{Computing methodologies~Machine learning}
\ccsdesc[500]{Computing methodologies~Natural language processing}
\ccsdesc[300]{Computing methodologies~Knowledge representation and reasoning}

\keywords{repository-level code generation, retrieval-augmented generation, code graph, feature-guided retrieval}

%% file: chapter/01-Introduction.tex
\section{Introduction}

Recent advances in large language models (LLMs) have shifted code generation research from isolated function-level completion tasks~\cite{yu-etal-2025-humaneval,austin2021programsynthesislargelanguage} toward repository-level generation ~\cite{Deveval,EvoCodeBench,CoderEval,CrossCodeEval}. 
Repository-level code generation requires an LLM not only to locate the code to be modified, but also to track its call relations, code dependencies, data constraints, and cross-file semantic links throughout the repository. The absence of such global relational context often leads to partial modifications, logic bugs, and violations of the system's existing architecture.
Consequently, capturing repository-wide code dependencies is the key to improving repository-level code generation~\cite{liu-2026-Beyond,LLM-Hallucinations,leanh2026treatcodenaturallanguage}.

Existing work mainly follows two directions. 
The first direction retrieves relevant code directly from repositories.
Representative approaches either embed source code for similarity-based retrieval~\cite{zhang-etal-2023-repocoder} or build structured representations, such as code context graphs and call-chain-aware graphs, to capture relationships among program entities~\cite{GraphCoder,RepoScope,GRACE}. They then retrieve candidate context through embedding search, graph expansion, or structure-aware reranking.
The second direction adopts agent-driven exploration. Starting from the task description, LLM agents iteratively inspect code, invoke tools, and, when available, leverage execution or test feedback to discover relevant implementation contexts~\cite{SWE-Agent,AutoCodeRover,RepairAgent}. 
Both directions substantially improve repository-level code generation by providing richer contextual information than isolated code completion.

Despite these advances, connecting functional requirements to dispersed implementation dependencies remains challenging. 
Code that is semantically similar to a requirement may omit supporting interfaces, while repository structure alone does not determine which dependencies are needed for implementation. Work on graph-based RAG has highlighted that pre-built structures may not align with the reasoning needs of individual queries~\cite{Chen2026}. 
Agent-driven exploration offers a flexible alternative, but incurs online interaction costs through multi-step LLM reasoning~\cite{SweRank}. Errors can also propagate across successive reasoning and tool-use steps, potentially derailing the exploration~\cite{Agentless}. 
These observations motivate an approach that uses functional organization to identify dependency candidates and combines semantic and structural heuristics to construct task-specific dependency context for implementation, without online LLM exploration.

\begin{figure*}[t]
    \centering
    \includegraphics[width=0.98\textwidth]{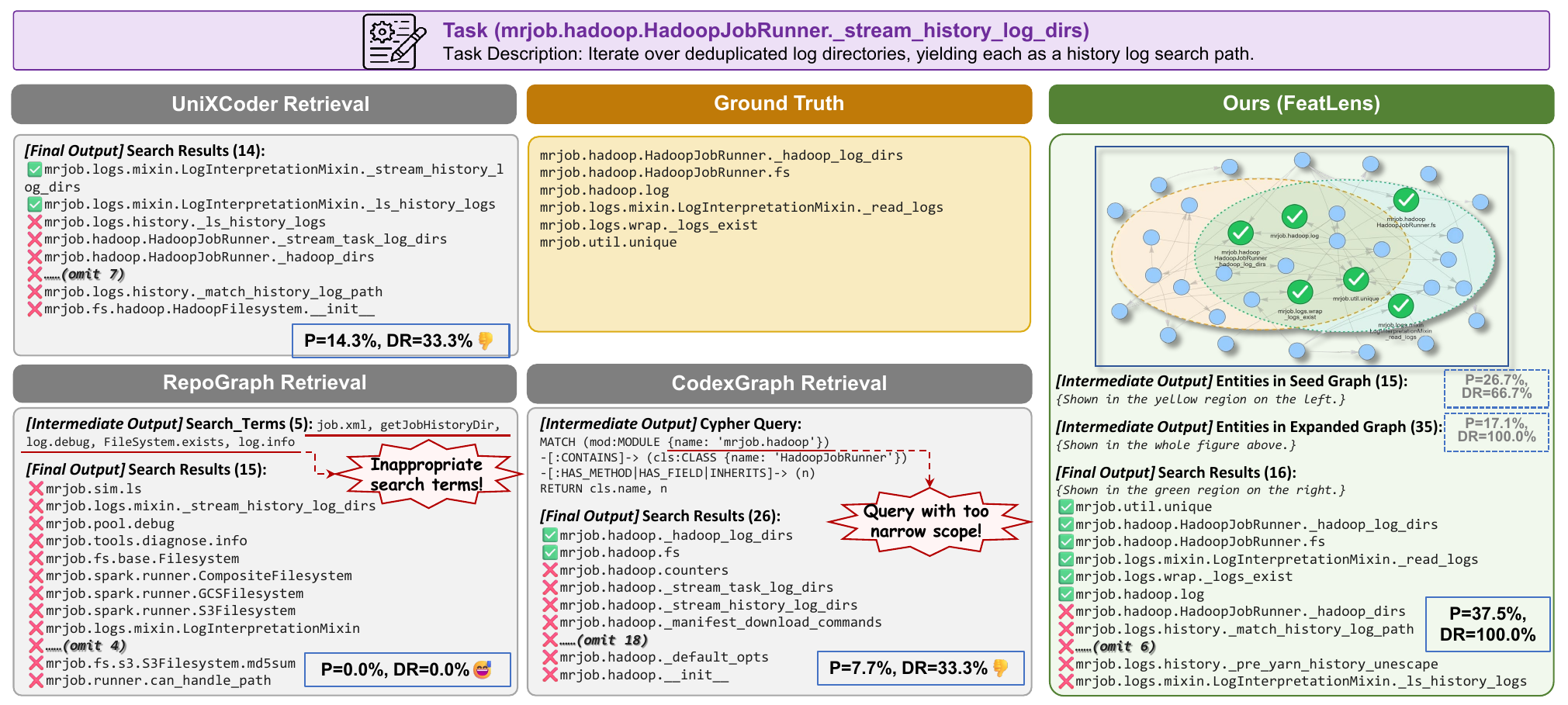}
    \caption{A motivating example of dependency retrieval for a history-log directory task in Mrjob. P and DR denote precision and dependency recall, respectively.}
    \Description{Comparison of UniXcoder, RepoGraph, CodexGraph, and FeatLens for a Mrjob history-log directory task. The panels list retrieved entities and mark matches to the reference dependencies. FeatLens additionally shows the seed, expanded, and final dependency contexts, connecting log-directory functions to cross-file helpers for deduplication, filesystem access, and existence checks.}
    \label{fig:motivation}
\end{figure*}

To this end, this paper proposes FeatLens, a feature-guided dependency retrieval approach for repository-level code generation.
The basic idea of FeatLens is to use features as a bridge connecting developers' task requirements to the target code and its dependencies. It establishes a bidirectional mapping between functional features and the repository code.
Specifically, FeatLens first produces a hierarchical functional summary of the repository and constructs a systematic index that links natural-language feature descriptions to function-level code entities. 
Then, upon receiving a generation task, it automatically matches relevant features, performs dependency seeding, expands the dependency context, and refines it through semantic-structural scoring and personalized PageRank.
Finally, the refined dependency context is provided to the LLM to support code generation.

We evaluate dependency retrieval on DevEval~\cite{Deveval} and EvoCodeBench~\cite{EvoCodeBench} and downstream function generation on DevEval using two LLMs.
Compared with UniXcoder RAG, FeatLens improves DR@15 by 11.2 percentage points on DevEval and 17.4 percentage points on EvoCodeBench.
On DevEval, it achieves the highest reported Pass@1 among non-oracle methods, reaching 42.24\% with DeepSeek-V3.2 and 55.03\% with GPT-5-mini, while producing the shortest code under both models.
These results are obtained without online LLM retrieval calls.
Component analysis shows that our feature index and feature-heuristic reranking (FeatHeur) mechanism can improve dependency retrieval over other settings.

The main contributions of this paper are as follows:

\begin{itemize}
\item We propose FeatLens, a feature-guided dependency retrieval approach for repository-level code generation that uses a feature-to-function index to connect task requirements with repository dependencies and construct dependency context on demand.

\item We present a semantic--structural dependency context refinement approach that combines code semantics and seed-derived structural signals through personalized PageRank to select a bounded dependency context.

\item We conduct an empirical evaluation on DevEval and EvoCodeBench showing higher dependency recall than the evaluated baselines, together with the highest non-oracle Pass@1 and shortest generated completions on DevEval under both generation models, without online LLM retrieval calls.
\end{itemize}

%% file: chapter/02-BackgroundMotivation.tex
\section{Background and Motivation}

More recently, features have been treated as a useful unit for organizing and implementing user requirements.
For example, EvoDev~\cite{EvoDev} decomposes user requirements into features and coordinates their implementation through dependency-aware planning, enabling design knowledge and business logic to propagate across development iterations. 
FeatX~\cite{FeatX} takes a feature-editing perspective for repository-level code evolution by translating feature edits into code patches.
These studies indicate that feature-level abstraction can provide useful guidance for repository-level development beyond isolated code fragments.

\subsection{Feature-Guided Dependency Retrieval}

Repository-level code generation tasks require a model to implement a target function within an existing repository. Such tasks are typically specified as functional requirements, yet correct implementations depend on task-relevant repository context, including existing functions, APIs, and data definitions. Following requirements-engineering terminology~\cite{agile}, we use \textit{feature} to denote the functional unit described by a requirement. Operationally, we represent a feature as an automatically identified cluster of structurally and semantically related functions paired with a natural-language summary, using this representation as a code-derived approximation of a functional unit. These indexed features serve as retrieval anchors that connect natural-language task intent to reusable repository code.

Let $R$ denote a repository, $C_R$ denote its code entities, and $L_R$ denote the structural relations among these entities. Code entities include functions, classes, files, and variables. Structural relations include calls, definitions, uses, containment, and imports. We associate $R$ with a set of functional features $F_R=\{f_1,\dots,f_n\}$, where each feature $f_i$ has a textual description $d_i$ and is supported by a subset of code entities $C_i\subseteq C_R$.

Given a requirement $q$ and a target function location $t$, feature-guided dependency retrieval aims to identify a compact dependency context $D_{q,t}=(V_{q,t},E_{q,t})$, where $V_{q,t}\subseteq C_R$ contains reusable entities and $E_{q,t}\subseteq L_R$ preserves their structural constraints for generation. Unlike conventional code search, the desired output is not merely code that is textually similar to $q$, but the dependencies that should be reused or respected during implementation. This requires aligning $q$ with repository features, recovering structural dependencies that may be lexically indirect, and avoiding excessive irrelevant context. The following example illustrates why existing retrieval strategies struggle with this formulation.

\subsection{Motivation and Challenges}
\label{sec:motivation-challenges}

\autoref{fig:motivation} illustrates dependency retrieval for implementing \texttt{\_stream\_\allowbreak history\_\allowbreak log\_\allowbreak dirs} in Mrjob's \texttt{HadoopJobRunner}. The task is to yield existing history-log directories while respecting the log-reading policy and removing duplicates. Correct implementation depends not only on log-directory routines, but also on cross-file helpers such as \texttt{unique} for deduplication and \texttt{\_logs\_exist} for existence checks through the repository's filesystem interface. This example highlights two challenges.

\paragraph{Challenge 1: Bridging task intent and dispersed implementation dependencies.}
Requirements describe a functional goal, whereas the code supporting that goal is distributed across files and abstraction levels. Similarity-based retrieval may favor history-log parsers or similarly named methods over supporting helpers whose names do not express their role in log discovery. Likewise, graph exploration centered on \texttt{HadoopJobRunner} can locate relevant methods without recovering the cross-file implementations they reuse. The difficulty is therefore to connect task intent to the entities that jointly support the requested behavior, rather than to code that merely shares its topic or location. A feature-level representation offers a useful retrieval anchor: the log-discovery feature groups directory-enumeration routines with reusable helpers, making their functional association explicit beyond lexical similarity and class boundaries.

\paragraph{Challenge 2: Preserving task-critical dependencies under a bounded context budget.}
A relevant feature provides an entry point, not a complete dependency context. Structural expansion can recover additional supporting interfaces, but it also introduces code from neighboring workflows that the target task does not require. Restricting the context by semantic similarity alone risks discarding generic yet necessary helpers; retaining the entire neighborhood exposes the generator to unnecessary code. The challenge is to distinguish task-critical dependencies from broadly related candidates while balancing coverage and context size. This requires considering both semantic relevance and structural relationships, rather than treating either as sufficient evidence of usefulness.

These challenges motivate FeatLens's combination of feature-guided dependency seeding and semantic--structural context refinement. Section~\ref{sec:case-study} revisits this example to examine how the retrieved interfaces relate to generation outcomes, and contrasts it with a case where refinement removes a required dependency.

%% file: chapter/03-Approach.tex
\section{Approach}

\begin{figure*}[t]
\centering
\includegraphics[width=0.95\textwidth]{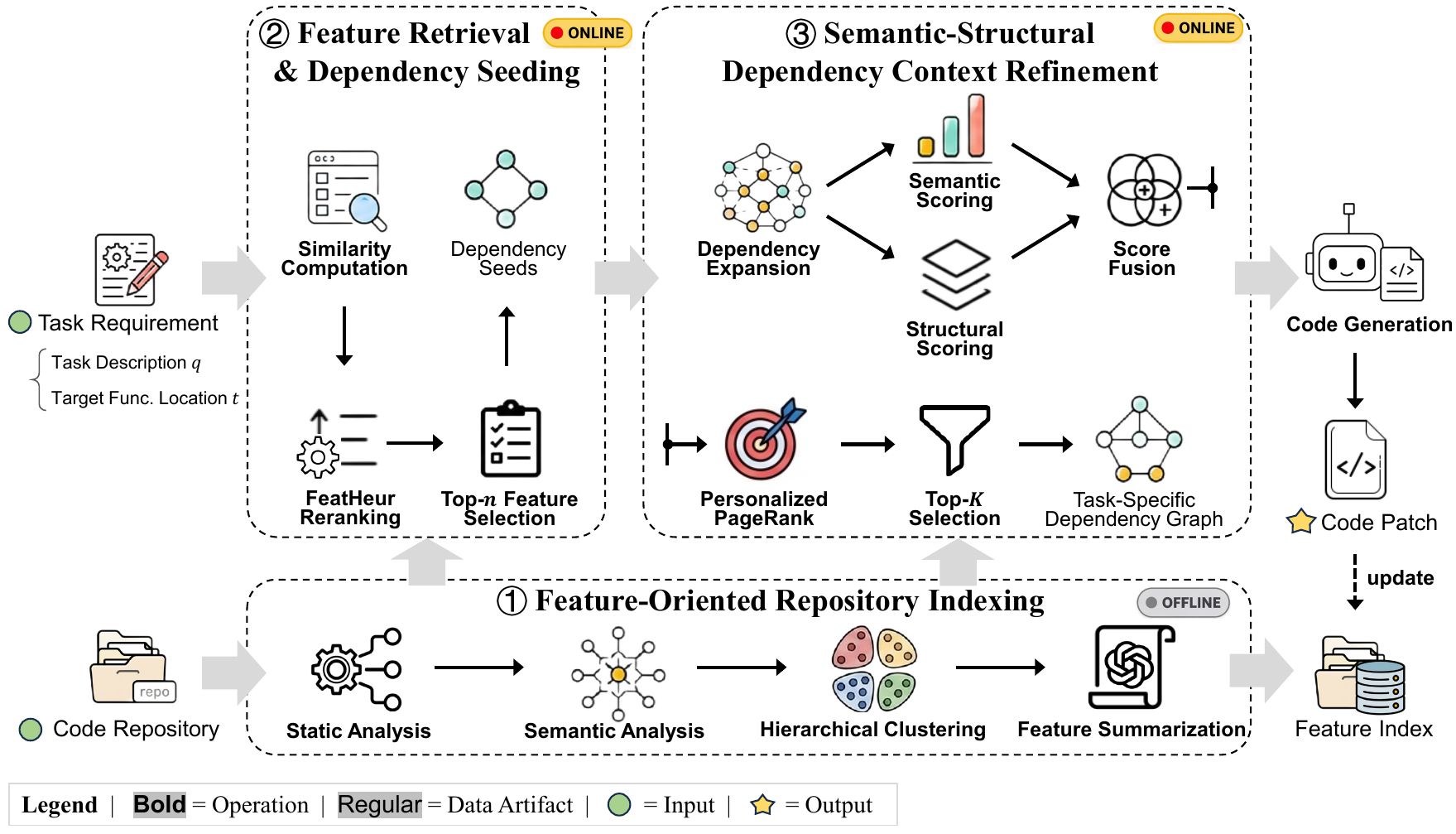}
\caption{Overview of FeatLens. An offline feature index guides online dependency seeding, expansion, and semantic--structural refinement to construct context for code generation.}
\Description{A three-stage workflow connects a code repository and a generation task to a code patch. Offline indexing combines static and semantic analysis, hierarchical clustering, and feature summarization. Online retrieval uses the task description and target location to select features and seed dependencies. Expansion adds related entities; semantic and structural scores guide personalized PageRank and Top-K selection. The resulting dependency graph supplies context to the code generator.}
\label{fig:framework}
\end{figure*}

\autoref{fig:framework} presents an overview of FeatLens, whose core method consists of three stages before code generation.
First, FeatLens builds a feature-oriented repository index from the source repository.
Second, given a task description and target function location, it performs feature retrieval and dependency seeding using the feature index.
Third, it performs dependency expansion and semantic-structural dependency context refinement, producing a task-specific dependency graph.
The resulting dependency graph provides context for downstream code generation.

\subsection{Feature-Oriented Repository Indexing}
\label{sec:Feature-Indexing}

The first stage builds an offline feature index that maps natural-language feature descriptions to function-level code entities.
FeatLens adapts the feature-oriented repository summarization procedure proposed by Zhu et al.~\cite{RepoSummary}, but uses the resulting feature structure as a retrieval index rather than as repository documentation.

Given a repository $R$, FeatLens first analyzes the repository from both structural and semantic perspectives.
Structural static analysis extracts code entities (functions, classes, and files) and records lightweight structural relations (calls and imports) using ENRE.
Semantic analysis represents each function according to its implementation semantics using lightweight descriptions and embeddings.
These structural relations capture coarse interactions among entities, while the semantic representations capture what functionality they provide.

To organize repository functions into feature-level groups, FeatLens applies hierarchical clustering based on the structural and semantic information extracted above.
Specifically, it uses Leiden community detection to first group files and then cluster functions within each file group using fused semantic and structural similarities.
Each resulting function cluster is summarized by an LLM into a natural-language feature description and linked to its constituent functions, yielding the feature index:

\begin{equation}
\mathcal{I}_R=\{f_i\}_{i=1}^{M}=\{(d_i,C_i^{\mathrm{func}})\}_{i=1}^{M},
\label{eq:feature-index}
\end{equation}

where $f_i$ denotes the $i$-th feature, $d_i$ is its natural-language description, and $C_i^{\mathrm{func}}$ denotes the subset of code entities associated with this feature at the function granularity.
This feature-to-function index serves as the basis for feature-guided retrieval in the next stage.

\subsection{Feature Retrieval and Dependency Seeding}

The second stage retrieves relevant features from the offline index and constructs a task-specific seed graph from their associated functions.
Given a task description $q$, comprising the functional requirement and parameter descriptions, and the target function location $t$, FeatLens first retrieves feature clusters whose descriptions are semantically aligned with the task. For each indexed feature $f_i=(d_i,C_i^{\mathrm{func}})$, FeatLens computes:

\begin{equation}
\mathrm{sim}(q,d_i)=\cos\left(\phi(q),\phi(d_i)\right).
\label{eq:sim}
\end{equation}

Ranking features solely by description similarity may retrieve features that are textually close to the task but weakly connected to the target location.
FeatLens therefore applies feature-heuristic reranking (FeatHeur) during dependency seeding to prioritize the feature associated with the target function. 
The target function's namespace identifies the feature cluster containing it, denoted by $f_t$, in the stored index. When available, $f_t$ is ranked first, followed by the remaining features in descending order of semantic similarity.
Equivalently, clusters are ordered by the following descending lexicographic key:
\begin{equation}
\mathrm{key}(q,f_i,t)=\bigl(\mathbf{1}[f_i=f_t],\mathrm{sim}(q,d_i)\bigr).
\label{eq:rerank}
\end{equation}
Notably, FeatLens operates on a repository in which the target function body is replaced with a \texttt{pass} statement before feature indexing. Consequently, both index construction and subsequent feature lookup have access only to the target function's signature and placeholder body, not its original implementation.
After FeatHeur reranking, FeatLens selects the top $n=3$ feature clusters to form $\mathcal{F}_q$ and collects their function entities as the dependency seeds:

\begin{equation}
S_q=\bigcup_{f_i\in \mathcal{F}_q} C_i^{\mathrm{func}}.
\label{eq:candidates}
\end{equation}

FeatLens constructs a directed seed graph $G_0=(V_0,E_0)$ from the selected functions, with $V_0=S_q$.
It matches these functions to ENRE entities using normalized qualified names and forms $E_0$ from the structural relations connecting pairs of seed nodes.
This seed graph links the feature-guided dependency seeds through their structural relations and serves as the starting point for dependency expansion.

\subsection{Semantic--Structural Dependency Context Refinement}

FeatLens expands the seed graph $G_0$ to recover dependencies beyond the selected features. From each seed function, it follows outgoing \texttt{Call}, \texttt{Use}, and \texttt{Contain} relations to function entities and incoming \texttt{Call} relations to callers, then adds the defining classes of all retained functions through incoming \texttt{Define} relations. Let $N_{\mathrm{func}}(V_0)$ denote these function neighbors and $C_{\mathrm{owner}}(V)$ the defining classes of functions in $V$. The expanded graph $G_e=(V_e,E_e)$ is constructed as:
\begin{align}
V_f &= V_0\cup N_{\mathrm{func}}(V_0),\nonumber\\
V_e &= V_f\cup C_{\mathrm{owner}}(V_f),
\label{eq:expand}
\end{align}
where $E_e$ contains the seed edges and the edges added during expansion.

To prioritize task-relevant nodes in $G_e$, FeatLens combines semantic relevance with structural guidance from the seeds. The semantic score $s_{\mathrm{semantic}}(v)$ is the cosine similarity between code-retrieval embeddings of the query and node code, clipped at zero; function nodes use their stored code, and class nodes use skeleton code. Structural guidance uses three potentially overlapping seed indicators relative to the target: \emph{same-file} identifies functions in the target's source file, \emph{same-feature} identifies functions in its indexed feature cluster, and \emph{similar-function} identifies the three functions ranked highest by BM25-based function-signature matching. Each seed $u$ receives a weight $b(u)=\alpha I_{\mathrm{sameFile}}(u)+\beta I_{\mathrm{sameFeature}}(u)+\gamma I_{\mathrm{similarFunction}}(u)$, which contributes to its newly expanded neighbors. With $N_e(v)$ containing the predecessors and successors of $v$ in $G_e$, the accumulated structural score is:
\begin{equation}
s_{\mathrm{struct}}(v)=\mathbf{1}[v\notin V_0]\sum_{u\in V_0\cap N_e(v)}b(u).
\label{eq:structural-reward}
\end{equation}
The fused score $w(v)=s_{\mathrm{semantic}}(v)+\lambda s_{\mathrm{struct}}(v)$ balances semantic relevance and structural guidance. FeatLens normalizes these scores into a restart vector $s$, using a uniform vector when all scores are zero.

FeatLens then applies personalized PageRank to propagate these preferences over $G_e$. With $A$ denoting the column-normalized transition matrix and $\mu$ the restart probability, the ranking scores are updated as:
\begin{equation}
p^{(r+1)}=(1-\mu)A p^{(r)}+\mu s.
\label{eq:ppr}
\end{equation}
After convergence, FeatLens selects the Top-$K$ nodes excluding the target function $t$ and retains the edges among them, yielding the task-specific dependency graph:
\begin{align}
G^*&=(V^*,E^*),\quad V^*=\mathrm{TopK}_{v\ne t}(p), \nonumber\\
E^*&=\{(u,v)\in E_e\mid u,v\in V^*\}.
\label{eq:reasoning-graph}
\end{align}
Algorithm~\ref{alg:subgraph-selection} summarizes this process. The selected dependency context is provided to the generator together with the task description and target function location.

\begin{algorithm}[t]
\caption{Semantic-Structural Dependency Context Refinement}
\label{alg:subgraph-selection}
\KwIn{Task $q$, seed graph $G_0$, target function $t$, hyperparameters $\Theta$.}
\KwOut{Task-specific dependency graph $G^*$.}
\SetKwFunction{RefineContext}{RefineContext}
\SetKwProg{Fn}{Function}{:}{}
\Fn{\RefineContext{$q,G_0,t,\Theta$}}{
Expand $G_0$ to obtain $G_e=(V_e,E_e)$ using Eq.~\eqref{eq:expand}\;
\ForEach{$v\in V_e$}{
    $s_{\mathrm{semantic}}(v)\leftarrow \max\{0,\cos(\phi_{\mathrm{code}}(q),\phi_{\mathrm{code}}(\mathrm{code}(v)))\}$\;
    Compute $s_{\mathrm{struct}}(v)$ using Eq.~\eqref{eq:structural-reward}\;
    $w(v)\leftarrow s_{\mathrm{semantic}}(v)+\lambda s_{\mathrm{struct}}(v)$\;
}
Normalize $w$ into $s$; use a uniform vector if all scores are zero\;
$p\leftarrow \mathrm{PersonalizedPageRank}(G_e,s,\mu)$\;
$V^*\leftarrow \mathrm{TopK}_{v\ne t}(p)$\;
$E^*\leftarrow \{(u,v)\in E_e\mid u,v\in V^*\}$\;
\Return{$G^*=(V^*,E^*)$}\;
}
\end{algorithm}

%% file: chapter/05-ExperimentalSettings.tex
\section{Evaluation}
\label{sec:evaluation}

We evaluate dependency retrieval quality and downstream generation utility; offline indexing investment, online LLM retrieval overhead, and dependency-context size; and the contributions of feature guidance, dependency seeding, and context refinement. Our evaluation is guided by the following research questions (RQs).

\begin{itemize}[leftmargin=*,widest={\textbf{RQ3}}]
    \item[\textbf{RQ1}] \textbf{Effectiveness of FeatLens.} Compared with alternative methods, how effective is FeatLens in retrieving task-relevant repository dependencies and generating code?

    \item[\textbf{RQ2}] \textbf{Indexing and Retrieval Overhead.} What offline indexing investment and online retrieval overhead does FeatLens incur, and how large is the resulting dependency context?

    \item[\textbf{RQ3}] \textbf{Component Contributions and Context-Budget Sensitivity.} How does refinement improve different retrieval strategies, how do Feature and FeatHeur contribute to retrieval quality, and how does the context budget affect dependency recall and precision?
\end{itemize}

\subsection{Experimental Setup}
\label{sec:experimental-setup}

\subsubsection{Benchmarks}

We evaluate FeatLens on DevEval~\cite{Deveval} and EvoCodeBench~\cite{EvoCodeBench}. Each task requires implementing a target function at a known repository location from its signature and natural-language requirement. Table~\ref{tab:dataset-statistics} summarizes the evaluated subsets. The complete retained task lists are available in our replication package.

\textbf{DevEval} contains 1,825 tasks from 115 Python projects across ten domains. Due to environment-reproduction difficulties, we retained 90 projects whose reference implementations passed the benchmark tests in our environment. We evaluate code generation on all 1,430 retained tasks and dependency retrieval on the 1,146 tasks with nonempty reference-dependency annotations.

\textbf{EvoCodeBench} contains 275 tasks from 25 Python projects. Environment-reproduction difficulties limited our evaluation to dependency retrieval. Three authors independently selected projects based on substantial task counts and high task density, source-code completeness and dependency traceability, and coverage of machine learning, computer vision, and LLM tooling, which are less represented in DevEval. The intersection of their selections yielded five projects with 123 tasks; the 109 tasks with nonempty reference-dependency annotations are used for retrieval evaluation.

\input{table/dataset-statistics}

For each task, we replace the target function body with a \texttt{pass} statement and then build an offline feature index from the resulting repository snapshot, as detailed in Section~\ref{sec:Feature-Indexing}.
RQ2 reports the mean offline index construction time and cost per task.

\subsubsection{Baseline Methods}

We compare FeatLens with seven baselines covering no-context generation, lexical and dense retrieval, repository-graph retrieval, multi-view structural context, and agentic repository exploration.

(1) \textbf{No-context}: the generator receives only the task requirement and target function signature, without retrieved repository context.

(2) \textbf{BM25 RAG}~\cite{BM25}: a lexical retrieval baseline that ranks repository functions by BM25 similarity between the task requirement and function code.

(3) \textbf{UniXcoder RAG}~\cite{unixcoder}: a dense retrieval baseline that embeds the requirement and repository functions with UniXcoder and ranks functions by embedding similarity.

(4) \textbf{RepoGraph}~\cite{RepoGraph}: a graph-based retrieval baseline that uses a line-level repository graph. An LLM generates search terms from the requirement, and the graph is searched to retrieve task-relevant context.

(5) \textbf{RepoScope}~\cite{RepoScope}: a static-analysis-based method that constructs a repository structural-semantic graph and assembles multi-view context, including callers, callees, similar functions, and similar code fragments, while preserving structural relations in the prompt.

(6) \textbf{LocAgent}~\cite{LocAgent}: an agent-based repository exploration method adapted for dependency localization. It iteratively searches repository structure and code through tool interactions to identify task-relevant locations. 

(7) \textbf{CodexGraph}~\cite{CodexGraph}: an LLM-based repository exploration method that stores repository entities and relations in Neo4j and uses LLM-generated Cypher queries to retrieve task-relevant context.

We additionally include an \textbf{Oracle} condition in which the code of the dependencies annotated in DevEval is supplied directly to the generation model as context, without retrieval.

For RepoGraph, CodexGraph, and LocAgent, online retrieval or exploration calls use DeepSeek-V3.2 with temperature 0. LocAgent is limited to five model calls, with a 32K-token context limit per call. CodexGraph uses the best setting of two retrieval iterations, with a one-iteration fallback for 101 of 1,430 tasks.

\subsubsection{Evaluation Metrics}

We evaluate both dependency retrieval and downstream generation.

(1) \textbf{DR@K} measures the fraction of reference dependencies covered by the Top-$K$ retrieved results:

\begin{equation}
\mathrm{DR}@K=\frac{|D_{\mathrm{retrieved}}\cap D_{\mathrm{gt}}|}{|D_{\mathrm{gt}}|}.
\label{eq:dr}
\end{equation}

Here, $D_{\mathrm{gt}}$ is the deduplicated union of the benchmark's intra-class, intra-file, and cross-file annotations.

(2) \textbf{Precision} (P) measures the proportion of retrieved results that are ground-truth dependencies.

(3) \textbf{F1} is the harmonic mean of $\mathrm{DR}@K$ and Precision, capturing the trade-off between dependency coverage and context noise.

(4) \textbf{DIR} (Dependency Invocation Rate)~\cite{hai-etal-2025-impacts} measures how many ground-truth dependencies are invoked by generated code:

\begin{equation}
\mathrm{DIR}=\frac{|D_{\mathrm{invoked}}\cap D_{\mathrm{gt}}|}{|D_{\mathrm{gt}}|}.
\label{eq:dir}
\end{equation}

Tasks with empty reference-dependency annotations are excluded from DIR computation.

(5) \textbf{Pass@1} measures the fraction of tasks whose single generated program passes all unit tests.

(6) \textbf{Code Length} measures the number of tokens obtained by tokenizing the generated code completion. We report the mean token count across evaluated completions.

(7) \textbf{CC} (Cyclomatic Complexity) is computed as an AST-based approximation of McCabe’s metric~\cite{McCabe1976}. Lower values indicate less structurally complex implementations.

\subsubsection{Implementation and Parameter Settings}

We use \texttt{CodeT5-base-multi-sum}~\cite{CodeT5} to generate function descriptions and Sentence-Transformers encoders~\cite{Sentence-BERT}, including \texttt{all-mpnet-base-v2} for clustering and \texttt{all-MiniLM-L6-v2} for feature retrieval. Node semantic scoring uses \texttt{BAAI/bge-code-v1}~\cite{BGE}. 
For context refinement, we set $\alpha=\beta=\gamma=0.1$ and $\lambda=1$. Personalized PageRank uses a restart probability of $\mu=0.15$, a maximum of 100 iterations, and a convergence tolerance of $10^{-6}$. 
For FeatLens, we evaluate final context budgets $K\in\{10,15,20\}$ and use $K=15$ for downstream generation. Code generation uses \texttt{deepseek-v3.2-251201} and \texttt{gpt-5-mini-2025-08-07} with temperature 0 and top-$p$ 0.95. DeepSeek-V3.2 runs in non-thinking mode, while GPT-5-mini uses the \texttt{minimal} reasoning-effort setting.
Retrieval queries consist of the task requirements and parameter descriptions and are constructed independently of the feature-index descriptions.
The LLM prompt templates used in our experiments are provided in our replication package.

\paragraph{Statistical analysis.}
On DevEval, we use paired task-level comparisons with repository-cluster bootstrap confidence intervals, two-sided sign-flip tests, and Holm correction for multiple comparisons. In Tables~\ref{tab:dependency-recall} and~\ref{tab:generation-results}, $^*$ marks a significant difference from FeatLens ($p_{\mathrm{adj}}<0.05$), in either direction. Statistical analyses cover DR, DIR, and Pass@1 on DevEval; Oracle is excluded. No statistical significance tests are performed on EvoCodeBench.

%% file: table/dataset-statistics.tex
\begin{table*}[t]
\centering
\caption{Statistics of the DevEval and EvoCodeBench subsets retained for evaluation.}
\label{tab:dataset-statistics}
\footnotesize
\setlength{\tabcolsep}{3pt}
\begin{tabular}{llcccccc}
\toprule
Dataset & Domain & \#Repos. & Avg. \#Files & Avg. \#Funcs. & Avg. LOC & \#Tasks & Avg. \#Deps./Task \\
\midrule
\multirow{11}{*}{DevEval} & Communication & 10 & 87.6 & 1485.2 & 31168.7 & 159 & 3.08 \\
& Database & 11 & 24.2 & 302.1 & 7206.6 & 153 & 3.75 \\
& Internet & 9 & 92.7 & 1039.3 & 23186.4 & 269 & 2.85 \\
& Multimedia & 4 & 35.3 & 340.8 & 6769.3 & 115 & 2.91 \\
& Scientific-Engineering & 6 & 114.2 & 954.7 & 34276.7 & 51 & 2.65 \\
& Security & 13 & 71.1 & 687.1 & 16519.9 & 161 & 2.89 \\
& Software-Development & 7 & 189.0 & 1147.3 & 69154.6 & 74 & 1.91 \\
& System & 9 & 41.8 & 371.1 & 8616.9 & 188 & 2.11 \\
& Text-Processing & 10 & 32.4 & 228.7 & 5653.1 & 47 & 1.81 \\
& Utilities & 11 & 57.7 & 444.4 & 14564.7 & 213 & 2.32 \\
& \textbf{Overall} & \textbf{90} & \textbf{70.9} & \textbf{711.0} & \textbf{20283.5} & \textbf{1430} & \textbf{2.72} \\
\midrule
EvoCodeBench & \textbf{Overall} & \textbf{5} & \textbf{81.0} & \textbf{768.0} & \textbf{17353.4} & \textbf{123} & \textbf{3.67} \\
\bottomrule
\end{tabular}
\end{table*}

%% file: chapter/06-ExperimentalResults.tex
\subsection{RQ1: Effectiveness of Dependency Retrieval and Code Generation}
\label{sec:rq1-retrieval}

\subsubsection{Dependency Retrieval Effectiveness}

As shown in \autoref{tab:dependency-recall}, FeatLens achieves the highest dependency recall at all three fixed retrieval budgets on both datasets. At $K=15$, which is also used for downstream generation, FeatLens reaches 0.527 on DevEval and 0.529 on EvoCodeBench, exceeding UniXcoder RAG by 11.2 and 17.4 percentage points, respectively. These results show that FeatLens covers more reference dependencies within the same retrieval budget. On DevEval, the recall gains over all fixed-budget baselines are statistically significant at each budget.

\input{table/dependency-recall}

The recall advantage also holds when FeatLens is given a smaller retrieval budget than competing methods. Its DR@10 exceeds the highest DR@20 among the fixed-budget baselines on both datasets: 0.498 versus 0.453 for UniXcoder RAG on DevEval, and 0.510 versus 0.402 for CodexGraph on EvoCodeBench. Thus, FeatLens covers more reference dependencies with half the maximum number of retrieved entries in these comparisons, showing that its recall advantage does not rely on a larger retrieval budget.

\subsubsection{Code Generation Performance}

\autoref{tab:generation-results} reports code generation performance on DevEval using the Top-15 retrieved context, except for LocAgent, whose variable-sized context is used instead.
FeatLens achieves the highest non-oracle Pass@1 under both DeepSeek-V3.2 (42.24\%) and GPT-5-mini (55.03\%). The gains over No-context, BM25 RAG, RepoGraph, and CodexGraph are statistically significant under both models. Under DeepSeek-V3.2, FeatLens achieves a Pass@1 of 42.24\%, compared with 41.40\% for UniXcoder RAG and 40.67\% for RepoScope, although neither difference is statistically significant.
These results support the downstream utility of the retrieved dependency context of FeatLens, with improvements over several baselines and competitive performance against the strongest alternatives.

\input{table/generation-results}

To examine whether these generation results are accompanied by greater reuse of repository dependencies, we compare dependency invocation rates (DIR). 
FeatLens increases DIR over No-context from 17.08\% to 52.91\% with DeepSeek-V3.2 and from 15.92\% to 53.58\% with GPT-5-mini, ranking first and second among non-oracle methods, respectively. Together with the Pass@1 results, these increases support the utility of the retrieved context for dependency reuse and function implementation. 
The Oracle condition, which directly supplies the benchmark-annotated reference dependencies, further raises DIR to 76.68\% and 81.02\%, yet its Pass@1 values of 42.73\% and 54.41\% remain within one percentage point of FeatLens's. This suggests that FeatLens's retrieved context may already provide much of the dependency information that the evaluated models can effectively exploit for correct implementation, leaving limited additional correctness gains from directly supplying the reference dependencies. 
FeatLens also produces relatively concise code with low control-flow complexity: code length is among the lowest under both models, while CC ranks second-lowest with DeepSeek-V3.2 and lowest with GPT-5-mini among non-oracle methods.

\begin{summarybox}
\textbf{Summary for RQ1:} FeatLens retrieves more reference dependencies within a limited retrieval budget: its Top-10 recall exceeds the Top-20 recall of every fixed-budget baseline on both datasets. The retrieved context also supports effective code generation, yielding the highest non-oracle Pass@1 point estimates under both models and results within one percentage point of the reference-dependency Oracle.
\end{summarybox}

\subsection{RQ2: Indexing and Retrieval Overhead}
\label{sec:rq2-cost}

\subsubsection{Offline Feature Indexing}

As shown in \autoref{tab:index-statistics}, the offline index organizes an average of 711.0 functions into 132.7 features on DevEval and 768.0 functions into 179.6 features on EvoCodeBench. Each feature contains about five functions on average, providing function groups for subsequent task-specific dependency seeding. 
Under our evaluation protocol, indexing is performed on each task's repository snapshot after masking the target function body. Mean construction time is 269.4 seconds on DevEval and 316.6 seconds on EvoCodeBench, with corresponding costs of USD 0.21 and USD 0.30 per task. These results indicate a modest offline indexing investment to establish the feature-to-function mappings used by online retrieval.

\input{table/index-statistics}

\subsubsection{Task-Specific Retrieval and Context Overhead}

\autoref{tab:engineering-overhead} reports graph sizes and online LLM retrieval overhead on DevEval. Repository-level graph sizes are amortized over tasks sharing each graph, whereas FeatLens reports its final task-specific graph after refinement. RepoGraph uses line-level nodes, while the other reported graphs use code entities. Retrieval tokens and calls count online LLM interactions; context tokens measure only the code supplied to the generator, using the \texttt{p50k\_base} tokenizer.

FeatLens supports the retrieval and generation effectiveness reported in RQ1 with a compact task-specific dependency graph. Its final graph averages 14.7 nodes and 13.7 edges, compared with the amortized repository-graph sizes of 95.6 nodes and 97.3 edges for RepoScope and 111.8 nodes and 243.6 edges for CodexGraph. The graph processed during FeatLens's refinement is also task-local, containing 35.4 nodes on average after expansion from feature-selected seeds (\autoref{tab:stage-retrieval}). Together with RQ1, these results show that feature-guided retrieval can provide effective dependency context through localized graph construction and refinement, without constructing a full repository dependency graph for each task.

This feature-guided retrieval and refinement pipeline requires no online LLM decisions. Like RepoScope, FeatLens incurs zero online LLM retrieval calls or tokens. In contrast, RepoGraph uses one model call to generate search terms, while CodexGraph and LocAgent rely on iterative LLM interactions, averaging 1.36 and 4.75 calls and approximately 12.1K and 31.0K retrieval tokens per task, respectively. FeatLens achieves the retrieval and generation results reported in RQ1 through feature-guided seeding and task-specific refinement, without the online model interactions used by LLM-assisted exploration.

\input{table/engineering-overhead}

The resulting code context averages 3217.2 tokens per task, close to LocAgent's 3457.8 tokens and approximately one-third of CodexGraph's 9694.4 tokens. RepoGraph and RepoScope supply shorter contexts of 2182.4 and 2097.4 tokens, respectively, but achieve lower dependency recall in RQ1. FeatLens therefore combines higher dependency coverage with a code-context size in the middle of the range reported for the evaluated methods, rather than relying on the largest context input to obtain its retrieval and generation results.

\begin{summarybox}
\textbf{Summary for RQ2:} With a modest offline indexing investment, FeatLens performs online dependency retrieval without LLM calls. The resulting task-specific context averages 14.7 nodes, 13.7 edges, and approximately 3.2K code tokens per task, supporting effective dependency retrieval and code generation with a compact graph and moderate-sized code context.
\end{summarybox}

\subsection{RQ3: Component Contributions and Context-Budget Sensitivity}
\label{sec:rq3-components}

RQ3 examines whether refinement improves different retrieval strategies, how Feature and FeatHeur contribute to dependency retrieval, and how the context budget affects the resulting coverage and precision.

\subsubsection{Refinement across Retrieval Strategies and Feature Guidance}

\input{table/initial-retrieval}

\autoref{tab:initial-retrieval} supports two complementary comparisons: changes from initial retrieval to the refined context within each row, and differences among retrieval strategies within each column. 
\texttt{Direct} retrieves the top 15 functions. For function-level methods, \texttt{Seed} matches the average count of functions in FeatLens's top three retrieved features. 
In contrast, feature-based methods collect the functions associated with their selected features.
\texttt{Final} applies the refinement stage, including structural expansion, scoring, and selection, with $K=15$.

Across all configurations, refinement improves DR, precision, and F1 over the seeds. For methods with Direct results, the refined context also outperforms direct retrieval at the same final budget of 15 entries. For example, DR increases from 0.309 to 0.398 for BM25-Code, from 0.425 to 0.480 for UniXcoder, and from 0.330 to 0.423 for Summary Only. These improvements show that refinement is not restricted to feature-based seeds: it also turns candidates from lexical, dense, and function-description retrieval into more effective dependency contexts.

Comparing the representations used for seeding shows the contribution of the Feature Index. Summary Only retrieves individual functions through their CodeT5-generated descriptions, whereas Feature further groups functions through clustering and retrieves the resulting feature summaries. 
Moving from Summary Only to Feature raises Seed DR from 0.349 to 0.390 and Seed F1 from 0.099 to 0.123. 
After refinement, although the DR gain is smaller (from 0.423 to 0.426), precision and F1 increase from 0.086 to 0.092 and from 0.133 to 0.141, respectively. 
Feature organization therefore improves the initial candidate set and retains an advantage in the concentration of reference dependencies in the final context.

Feature Only and BM25-Signature yield similar Final DR (0.426 and 0.425), but combining Feature with FeatHeur raises it to 0.527. 
Actually, BM25-Signature in the table serves as a function-level heuristic counterpart: it uses signature matching to select candidates, whereas FeatHeur in FeatLens prioritizes the feature group containing the target function during dependency seeding. 
Viewed from the Feature configuration, adding FeatHeur improves Seed/Final DR from 0.390/0.426 to 0.491/0.527. Viewed from the signature-based heuristic configuration, the feature-guided component improves these values from 0.358/0.425 to 0.491/0.527. The complete FeatLens approach also achieves the highest precision and F1 at both Seed and Final. 
Together, these comparisons support the complementary value of functional grouping and heuristic guidance rather than relying on either retrieval signal alone.

\subsubsection{Context Refinement and Context-Budget Sensitivity}

\input{table/deveval-stages}

\autoref{tab:stage-retrieval} traces how refinement expands the candidate set and then selects a bounded context. On DevEval, expansion increases the average node count from 17.8 to 35.4 and DR from 0.491 to 0.655; on EvoCodeBench, the corresponding changes are 15.0 to 30.5 nodes and 0.478 to 0.671 DR. Subsequent ranking and selection at $K=15$ retain DR of 0.527 and 0.529, respectively. Although selection discards some dependencies from the expanded candidates, the final context improves aggregate DR, precision, and F1 over the seeds on both datasets. Refinement thus preserves part of the coverage gained through expansion while producing a more concentrated dependency context under the output budget.

The final context budget controls the trade-off between dependency recall and retrieval precision. On both datasets, $K=10$ yields the highest aggregate precision and F1, whereas $K=20$ yields the highest recall (0.547 on DevEval and 0.553 on EvoCodeBench). The benefit of additional entries varies across repositories: increasing $K$ from 15 to 20 raises recall on EasyVolcap but not on ollama-python. We use $K=15$ as a common operating budget for generation: it retains more than 95\% of the recall obtained at $K=20$ on both datasets while reducing the maximum node budget by 25\%. This provides a practical balance between dependency coverage and context size among the evaluated budgets.

\begin{summarybox}
\textbf{Summary for RQ3:} Refinement improves dependency retrieval across all evaluated seeding strategies, not only feature-based retrieval. Feature organization and FeatHeur provide complementary guidance, with their combination achieving the strongest Seed and Final results. The context budget controls the trade-off between dependency coverage and concentration; $K=15$ provides a practical operating compromise.
\end{summarybox}

\subsection{Case Study}
\label{sec:case-study}

\begin{figure*}[t]
    \centering
    \includegraphics[width=0.98\textwidth]{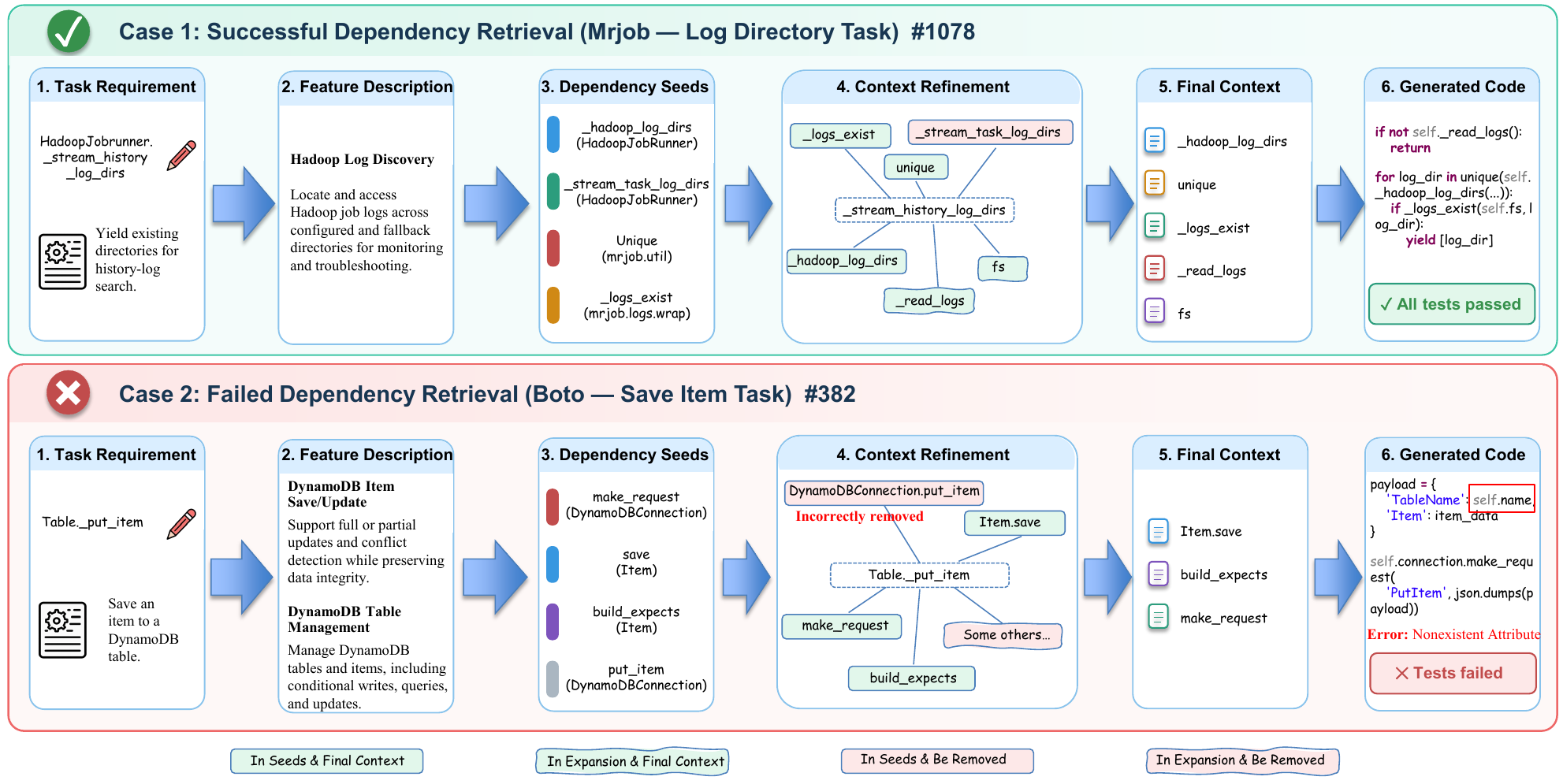}
    \caption{Case studies of FeatLens dependency retrieval and code generation.}
    \Description{Two workflows trace task requirements through feature retrieval, dependency seeding, context refinement, and code generation. In the Mrjob example, the final context retains directory-enumeration, deduplication, existence-checking, log-reading, and filesystem helpers; the generated code passes the tests. In the Boto example, the required DynamoDB write interface is present in the seeds but removed during refinement. The generated code instead constructs a request using a nonexistent table-name attribute and fails the tests.}
    \label{fig:case-study}
\end{figure*}

We use two contrasting cases to examine whether retrieved context preserves the interfaces needed for correct implementation (\autoref{fig:case-study}). The Mrjob case illustrates the benefit of recovering cross-file dependencies beyond the target class; the Boto case exposes the risk of removing a required interface during refinement. Generation results use GPT-5-mini.

\paragraph{Success: preserving repository-specific interface behavior.}
The Mrjob task from Section~\ref{sec:motivation-challenges} distinguishes locating relevant code from retrieving the interfaces that constrain a correct implementation. In particular, \texttt{\_logs\_exist} checks paths through the repository's filesystem abstraction and returns \texttt{None} on an \texttt{IOError}. Its implementation therefore provides concrete guidance on filesystem access and error handling that a log-related method name or target-class location alone does not supply.

LocAgent and CodexGraph retrieve context around \texttt{HadoopJobRunner}, but neither supplies the implementation of \texttt{\_logs\_exist} in its final generation context. Their generated code violates the corresponding behavior in different ways: LocAgent uses \texttt{os.path.exists} and yields no directory in the HDFS-path test, whereas CodexGraph omits the existence check and fails to skip inaccessible paths. BM25 and UniXcoder exhibit a related path-semantics error, appending an unnecessary \texttt{history/} suffix to the candidate directory.

FeatLens instead connects the log-discovery feature to cross-file helpers and expands the seeds to include the log-reading and filesystem interfaces. Refinement retains these dependencies in the final context. With this context, GPT-5-mini reuses the repository interfaces, yields the candidate directory without altering its path, and passes the tests. The contrast illustrates the value of retaining behavior-defining dependencies rather than merely retrieving code near the target or about the same topic. 

\paragraph{Failure: removing a critical interface despite a relevant feature match.}
The Boto task, \texttt{Table.\_put\_item}, exposes the complementary limitation: a useful feature-guided entry point does not guarantee that refinement preserves the interface needed for implementation. The selected feature covers the DynamoDB item-save workflow, and the initial seeds already contain \texttt{DynamoDBConnection.put\_item}. However, refinement removes this interface when selecting 15 nodes from the 42-node expanded context, while retaining more general request-construction routines such as \texttt{make\_request}.

The generated code constructs the request payload itself and accesses the nonexistent attribute \texttt{self.name}, causing an \texttt{AttributeError}. In contrast, the BM25 and UniXcoder implementations invoke \texttt{self.connection.put\_item(self.table\_name, ...)} and pass the tests. Here, the retrieval failure lies not in discovering the relevant workflow or admitting the correct interface into the candidate set, but in retaining that interface under the context budget.

Together, the cases distinguish functional relevance from implementation sufficiency: feature guidance can expose dependencies missed by topic- or location-centered retrieval, but budgeted refinement must preserve the interfaces that define the required behavior, and the generator must still use them correctly.

%% file: table/dependency-recall.tex
\providecommand{\overalldr}[2][\dagger]{\makebox[\dimexpr0.345\textwidth+4\tabcolsep\relax][c]{#2\rlap{$^{#1}$}}}
\begin{table*}[t]
\centering
\caption{Dependency recall of retrieval and exploration methods on DevEval and EvoCodeBench.}
\label{tab:dependency-recall}
\footnotesize
\setlength{\tabcolsep}{4.5pt}
\begin{tabular}{@{}p{0.16\textwidth}*{6}{>{\centering\arraybackslash}p{0.115\textwidth}}@{}}
\toprule
\multirow{2}{*}{Method} & \multicolumn{3}{c}{DevEval} & \multicolumn{3}{c}{EvoCodeBench} \\
\cmidrule(lr){2-4}\cmidrule(lr){5-7}
& DR@10 & DR@15 & DR@20 & DR@10 & DR@15 & DR@20 \\
\midrule

BM25 RAG & 0.261$^*$ & 0.296$^*$ & 0.322$^*$ & 0.204 & 0.241 & 0.273 \\
UniXcoder RAG & 0.354$^*$ & 0.415$^*$ & 0.453$^*$ & 0.311 & 0.355 & 0.371 \\

RepoGraph & 0.097$^*$ & 0.099$^*$ & 0.100$^*$ & 0.106 & 0.106 & 0.108 \\
RepoScope & 0.272$^*$ & 0.338$^*$ & 0.345$^*$ & 0.299 & 0.332 & 0.334 \\

LocAgent & \multicolumn{3}{c}{\overalldr[\dagger,*]{0.491}} & \multicolumn{3}{c}{\overalldr{0.477}} \\
CodexGraph & 0.239$^*$ & 0.309$^*$ & 0.344$^*$ & 0.244 & 0.354 & 0.402 \\

FeatLens & \textbf{0.498} & \textbf{0.527} & \textbf{0.547} & \textbf{0.510} & \textbf{0.529} & \textbf{0.553} \\
\bottomrule
\end{tabular}
\par\smallskip
\begin{minipage}{\linewidth}
\footnotesize\raggedright
$^\dagger$LocAgent returns variable-sized results; its recall differs significantly from FeatLens on DevEval at $K=15,20$.
\end{minipage}
\end{table*}

%% file: table/generation-results.tex
\begin{table*}[t]
\centering
\caption{Code generation performance on DevEval with DeepSeek-V3.2 and GPT-5-mini.}
\label{tab:generation-results}
\footnotesize
\setlength{\tabcolsep}{3pt}
\begin{tabular*}{\textwidth}{@{\extracolsep{\fill}} l *{8}{>{\centering\arraybackslash}p{0.085\textwidth}} @{}}
\toprule
\multirow{2}{*}{Method} & \multicolumn{4}{c}{DeepSeek-V3.2} & \multicolumn{4}{c}{GPT-5-mini} \\
\cmidrule(lr){2-5}\cmidrule(lr){6-9}
& DIR(\%) & Pass@1(\%) & Code Len. & CC & DIR(\%) & Pass@1(\%) & Code Len. & CC \\
\midrule

No-context & 17.08$^*$ & 17.20$^*$ & 65.8 & 4.05 & 15.92$^*$ & 29.23$^*$ & 252.2 & 16.00 \\

BM25 RAG & 42.59$^*$ & 32.17$^*$ & 66.1 & 3.95 & 44.07$^*$ & 48.18$^*$ & 128.6 & 7.82 \\
UniXcoder RAG & 49.18$^*$ & 41.40 & 65.8 & \textbf{3.82} & 51.92 & 53.43 & 108.8 & 6.55 \\

RepoGraph & 29.92$^*$ & 23.01$^*$ & 73.6 & 4.72 & 36.21$^*$ & 36.29$^*$ & 167.0 & 10.43 \\
RepoScope & 37.77 & 40.67 & 120.9 & 5.77 & 38.44 & 47.91 & 119.9 & 6.84 \\

LocAgent & 45.22$^*$ & 33.71$^*$ & 65.5 & 4.46 & \textbf{61.59}$^*$ & 50.98 & 115.5 & 7.13 \\
CodexGraph & 48.16 & 35.61$^*$ & 76.2 & 4.57 & 45.98$^*$ & 36.88$^*$ & 138.5 & 9.21 \\

FeatLens & \textbf{52.91} & \textbf{42.24} & \textbf{65.3} & 3.87 & 53.58 & \textbf{55.03} & \textbf{101.1} & \textbf{6.10} \\

\midrule

\textit{Oracle} & \textit{76.68} & \textit{42.73} & \textit{68.0} & \textit{4.06} & \textit{81.02} & \textit{54.41} & \textit{112.3} & \textit{7.03} \\

\bottomrule
\end{tabular*}
\end{table*}

%% file: table/index-statistics.tex
\begin{table}[t]
\centering
\caption{Feature-index statistics and mean offline construction time and cost per task. }
\label{tab:index-statistics}
\footnotesize
\setlength{\tabcolsep}{0pt}
\begin{tabular*}{\columnwidth}{@{\extracolsep{\fill}}lccccc@{}}
\toprule
Dataset & \#Func. & \#Feat. & \#Func./Feat. & Time (s) & Cost (USD) \\
\midrule
DevEval & 711.0 & 132.7 & 5.53 & 269.4 & 0.21 \\
EvoCodeBench & 768.0 & 179.6 & 5.25 & 316.6 & 0.30 \\
\bottomrule
\end{tabular*}
\end{table}

%% file: table/engineering-overhead.tex
\begin{table*}[t]
\centering
\caption{Engineering overhead of retrieval and exploration methods on DevEval (per-task averages).}
\label{tab:engineering-overhead}
\footnotesize
\setlength{\tabcolsep}{3pt}
\begin{tabular*}{\textwidth}{@{\extracolsep{\fill}} l c c c c c c @{}}
\toprule
\multirow{2}{*}{Method} & \multirow{2}{*}{\shortstack{Graph Scope +\\Retrieval Mode}} & \multicolumn{2}{c}{Graph Size} & \multicolumn{2}{c}{LLM Retrieval} & \multirow{2}{*}{\shortstack{Context Tokens}} \\
\cmidrule(lr){3-4}\cmidrule(lr){5-6}
& & \shortstack{\#Nodes} & \shortstack{\#Edges} & \shortstack{Retrieval Tokens} & \shortstack{Model Calls} & \\
\midrule
RepoGraph & Repo + LLM & 38.3 & 1580.0 & 287.2 & 1 & 2182.4 \\
RepoScope & Repo + Static & 95.6 & 97.3 & 0 & 0 & 2097.4 \\
CodexGraph & Repo + LLM & 111.8 & 243.6 & 12{,}097.1 & 1.36 & 9694.4 \\
LocAgent & None + LLM & -- & -- & 30{,}958.7 & 4.75 & 3457.8 \\
FeatLens & Task + Static & 14.7 & 13.7 & 0 & 0 & 3217.2 \\
\bottomrule
\end{tabular*}
\end{table*}

%% file: table/initial-retrieval.tex
\begin{table*}[t]
\centering
\caption{Dependency retrieval before and after refinement with different seeding strategies and feature guidance on DevEval.}
\label{tab:initial-retrieval}
\footnotesize
\setlength{\tabcolsep}{3pt}
\begin{tabular*}{\textwidth}{@{\extracolsep{\fill}} l *{9}{>{\centering\arraybackslash}p{0.05\textwidth}} @{}}
\toprule
\multirow{2}{*}{Method} & \multicolumn{3}{c}{Direct} & \multicolumn{3}{c}{Seed} & \multicolumn{3}{c}{Final} \\
\cmidrule(lr){2-4}\cmidrule(lr){5-7}\cmidrule(lr){8-10}
& DR & P & F1 & DR & P & F1 & DR & P & F1 \\
\midrule
BM25-Code & .309 & .061 & .094 & .324 & .058 & .090 & .398 & .081 & .124 \\
BM25-Description & .313 & .062 & .096 & .324 & .061 & .094 & .402 & .082 & .127 \\
BM25-Signature(aka. FeatHeur) & .343 & .068 & .106 & .358 & .066 & .102 & .425 & .087 & .133 \\
UniXcoder & .425 & .085 & .132 & .438 & .084 & .129 & .480 & .098 & .151 \\
Summary Only & .330 & .065 & .101 & .349 & .064 & .099 & .423 & .086 & .133 \\
Feature & / & / & / & .390 & .082 & .123 & .426 & .092 & .141 \\
\quad \textbf{FeatLens: Feature (+FeatHeur)} & / & / & / & \textbf{.491} & \textbf{.104} & \textbf{.157} & \textbf{.527} & \textbf{.112} & \textbf{.172} \\
\bottomrule
\end{tabular*}
\end{table*}

%% file: table/deveval-stages.tex
\begin{table*}[t]
\centering
\caption{Graph size and dependency retrieval performance across FeatLens stages and context budgets on DevEval and EvoCodeBench.}
\label{tab:stage-retrieval}
\footnotesize
\setlength{\tabcolsep}{0pt}
\begin{tabular}{@{}>{\raggedright\arraybackslash}p{0.17\textwidth}*{15}{>{\centering\arraybackslash}p{0.0553\textwidth}}@{}}
\toprule
\multirow{3}{*}{\shortstack[l]{Domain /\\Repository}} & \multicolumn{4}{c}{\multirow{2}{*}{Dependency Seeds}} & \multicolumn{2}{c}{\multirow{2}{*}{\shortstack{Dependency\\Expansion}}} & \multicolumn{9}{c}{Task-Specific Dependency Graph} \\
\cmidrule(lr){8-16}
& & & & & & & \multicolumn{3}{c}{K=10} & \multicolumn{3}{c}{K=15} & \multicolumn{3}{c}{K=20} \\
\cmidrule(lr){2-5}\cmidrule(lr){6-7}\cmidrule(lr){8-10}\cmidrule(lr){11-13}\cmidrule(lr){14-16}
& \#N. & DR & P & F1 & \#N. & DR & DR & P & F1 & DR & P & F1 & DR & P & F1 \\
\midrule
\multicolumn{16}{@{}c@{}}{\begingroup\setlength{\fboxsep}{0pt}\colorbox[HTML]{EFEFEF}{\makebox[\dimexpr0.9995\textwidth\relax][c]{\strut\textbf{DevEval}}}\endgroup} \\
\midrule
Communication & 17.9 & .444 & .097 & .145 & 36.8 & .588 & .435 & .143 & .200 & .483 & .109 & .168 & .505 & .091 & .146 \\
Database & 15.2 & .410 & .126 & .169 & 33.4 & .585 & .422 & .184 & .221 & .435 & .133 & .179 & .464 & .110 & .158 \\
Internet & 21.2 & .456 & .093 & .144 & 41.0 & .631 & .464 & .144 & .209 & .493 & .105 & .166 & .512 & .084 & .139 \\
Multimedia & 24.1 & .500 & .079 & .127 & 41.0 & .615 & .468 & .141 & .199 & .515 & .115 & .173 & .538 & .095 & .150 \\
Sci.-Eng. & 16.9 & .319 & .084 & .123 & 34.3 & .479 & .357 & .128 & .175 & .372 & .090 & .136 & .372 & .070 & .111 \\
Security & 16.9 & .522 & .112 & .167 & 32.8 & .717 & .542 & .166 & .228 & .565 & .119 & .179 & .583 & .096 & .152 \\
Softw.-Dev. & 15.5 & .439 & .082 & .124 & 34.1 & .599 & .460 & .117 & .173 & .474 & .083 & .133 & .474 & .067 & .110 \\
System & 14.4 & .530 & .104 & .161 & 33.7 & .744 & .573 & .137 & .206 & .616 & .104 & .167 & .631 & .083 & .139 \\
Text-Proc. & 15.5 & .532 & .092 & .147 & 28.0 & .587 & .504 & .120 & .183 & .549 & .095 & .155 & .549 & .075 & .126 \\
Utilities & 16.6 & .629 & .130 & .199 & 30.5 & .763 & .612 & .175 & .256 & .630 & .128 & .202 & .659 & .109 & .177 \\
\textbf{Overall} & \textbf{17.8} & \textbf{.491} & \textbf{.104} & \textbf{.157} & \textbf{35.4} & \textbf{.655} & \textbf{.498} & \textbf{.152} & \textbf{.213} & \textbf{.527} & \textbf{.112} & \textbf{.172} & \textbf{.547} & \textbf{.092} & \textbf{.147} \\
\midrule
\multicolumn{16}{@{}c@{}}{\begingroup\setlength{\fboxsep}{0pt}\colorbox[HTML]{EFEFEF}{\makebox[\dimexpr0.9995\textwidth\relax][c]{\strut\textbf{EvoCodeBench}}}\endgroup} \\
\midrule
litdata & 13.3 & .441 & .142 & .183 & 25.5 & .611 & .491 & .187 & .223 & .499 & .134 & .180 & .537 & .116 & .167 \\
UniRef & 13.2 & .513 & .152 & .210 & 24.6 & .700 & .511 & .196 & .231 & .530 & .149 & .199 & .555 & .134 & .192 \\
EasyVolcap & 14.4 & .388 & .092 & .124 & 42.8 & .730 & .467 & .111 & .158 & .467 & .078 & .118 & .481 & .067 & .104 \\
ollama-python & 15.6 & .567 & .085 & .140 & 25.0 & .733 & .567 & .125 & .190 & .650 & .106 & .173 & .650 & .079 & .135 \\
tanuki\_py & 31.0 & .712 & .129 & .202 & 55.1 & .763 & .656 & .300 & .385 & .684 & .217 & .311 & .684 & .163 & .250 \\
\textbf{Overall} & \textbf{15.0} & \textbf{.478} & \textbf{.128} & \textbf{.174} & \textbf{30.5} & \textbf{.671} & \textbf{.510} & \textbf{.176} & \textbf{.221} & \textbf{.529} & \textbf{.130} & \textbf{.181} & \textbf{.553} & \textbf{.110} & \textbf{.162} \\
\bottomrule
\end{tabular}
\par\smallskip
\begin{minipage}{\linewidth}
\footnotesize\raggedright
\textit{Notes.} Sci.-Eng. = Scientific-Engineering; Softw.-Dev. = Software-Development; Text-Proc. = Text-Processing.
\end{minipage}
\end{table*}

%% file: chapter/07-Discussion.tex
\section{Discussion}
\label{sec:discussion}

\subsection{Implications and Applicability Boundaries}

Overall, the results highlight the value of combining functional abstraction with repository structure to construct task-relevant dependency context. The feature index connects task requirements to groups of repository functions, while feature-heuristic reranking prioritizes the group containing the target function. Structural expansion and semantic--structural refinement then transform these groups into a bounded dependency context for code generation. The component results in Section~\ref{sec:rq3-components} support the effectiveness of feature-guided seeding with target-location guidance and show that subsequent expansion and refinement improve dependency retrieval over the initial seeds.

The results further suggest that feature guidance and structural reasoning play complementary roles. Feature-guided seeding provides functionally coherent entry points, while expansion recovers dependencies beyond the initial feature groups and refinement prioritizes them under a bounded context budget. This design provides a practical balance between dependency coverage and context concentration, and also motivates adaptive context budgets based on retrieval confidence or structural connectivity.

The two case studies further highlight the importance of preserving repository-specific interfaces during context refinement. In Mrjob, the final context retains the repository-specific interfaces reused by the passing implementation. In Boto, the relevant interface is identified in the seeds but is not retained in the final context, and the generated implementation instead constructs a request using a nonexistent attribute. These observations suggest that refinement can further benefit from explicitly considering which interfaces encode implementation-specific behavior. A direction for future work is to combine context refinement with execution feedback to further improve dependency selection and generated implementations.

FeatLens is most naturally applicable when the target function location is known and the required implementation depends on statically traceable repository entities, such as reusable functions, class members, and cross-file APIs. Its benefits are likely to be less predictable when behavior is introduced through runtime binding, implicit framework conventions, external configuration, or low-level interfaces that are weakly represented by feature clusters. Tasks requiring coordinated edits across multiple files or iterative test-driven repair also extend beyond the setting evaluated in this paper. These boundaries define the current role of FeatLens as a repository-aware dependency context constructor for function generation, while motivating extensions toward multi-file feature implementation and execution-guided refinement.

\subsection{Threats to Validity}
\label{sec:threats}

\textbf{External validity.} Our evidence comes from selected Python repositories in DevEval and EvoCodeBench. DevEval projects were retained when their reference implementations passed the tests in our configured environments, while EvoCodeBench projects were selected using explicit domain and dependency-traceability criteria rather than random sampling. EvoCodeBench supports retrieval conclusions only in this study. The findings therefore characterize repository-aware function generation and dependency retrieval in the evaluated Python settings; generalization to other languages, industrial repositories, dynamic dependency structures, and full feature-development benchmarks remains to be established.

\textbf{Construct validity.} DR, precision, and F1 measure retrieval coverage and context concentration; DIR measures whether generated code invokes annotated dependencies; and Pass@1 measures executable correctness. Code length and cyclomatic complexity provide proxies for conciseness and structural complexity, while the retrieval metrics inherit the scope and granularity of the benchmark annotations.

\textbf{Internal validity.} Results may depend on retrieval parameters, prompts, candidate budgets, and implementation choices across systems. The downstream generation comparison fixes $K=15$ without independent held-out tuning. Finally, offline feature summarization and static relations may miss runtime binding and implicit framework behavior.

%% file: chapter/08-RelatedWork.tex
\section{Related Work}

Related work on repository-level code generation can be broadly grouped into three directions: dependency retrieval, LLM-based repository exploration, and feature-oriented software development. FeatLens connects these lines of research by using feature-level intent to guide dependency context retrieval.

\subsection{Dependency Retrieval in Repository-Level Code Generation}

Retrieval-augmented code generation selects relevant repository context through semantic or structural signals. Semantic methods use sparse matching~\cite{BM25,CodeSearchisAllYouNeed,FT2Ra}, code embeddings~\cite{unixcoder,codebert}, or learned retrieval strategies such as RLCoder~\cite{RLCoder}, AlignCoder~\cite{AlignCoder}, and Hierarchical Embedding Fusion~\cite{HierarchicalEmbeddingFusion}. Since semantic similarity may miss cross-file dependencies~\cite{DRACO}, structure-aware methods further exploit code graphs. RepoGraph~\cite{RepoGraph} and GraphCoder~\cite{GraphCoder} model repository dependencies for graph-based retrieval, while later work considers multi-hop reasoning~\cite{RepoHyper}, multi-granularity structure~\cite{GRACE}, architecture preservation~\cite{RepoScope}, and knowledge graphs~\cite{KnowledgeGraphRepoGeneration}. InlineCoder~\cite{hu2026linecontextrepositorylevelcode} organizes dependency context around a generated draft, whereas FastCoder~\cite{FastCoder} emphasizes retrieval efficiency.

Overall, these methods organize retrieval mainly around repository code or intermediate generation states. FeatLens instead uses functional features as retrieval anchors, linking each feature to candidate functions that serve as seeds for dynamic dependency expansion and refinement.

\subsection{LLM-Based Repository Exploration}

LLM-based repository exploration lets agents iteratively search and inspect repositories to identify task-relevant context. 
Early agents, including SWE-Agent~\cite{SWE-Agent}, AutoCodeRover~\cite{AutoCodeRover}, and CodeAgent~\cite{zhang-etal-2024-codeagent}, explore repositories mainly through autonomous tool use.
This flexibility is useful, but an unconstrained search space makes reliable localization difficult.
Subsequent work introduces structured repository graphs to guide localization and reasoning~\cite{LocAgent, Jiang-2025-Issue}.
CodexGraph~\cite{CodexGraph} enables LLMs to explore a repository graph database through Cypher queries,
while GraphCodeAgent~\cite{GraphCodeAgent} couples a requirement graph with a code graph for agent-driven multi-hop reasoning between intent and dependencies.
More recent approaches pursue stronger but heavier exploration, using Monte Carlo tree search~\cite{LingmaAgent}, autonomous multi-round context construction~\cite{RepoMaster}, or knowledge graphs with working memory for long-horizon navigation~\cite{Prometheus}.

These methods improve contextual coverage but often require repeated LLM decisions. FeatLens instead narrows the search space with feature-guided dependency seeds and retrieves dependency context without online LLM decisions.

\subsection{Feature-Oriented Software Development}

Feature-oriented software development views features as high-level units for understanding, organizing, and evolving software functionality~\cite{Batory-2004-Feature-Oriented,Kästner2013}. 
Feature location techniques aim to identify the program entities responsible for implementing a given functionality~\cite{Dit-2013-FeatureLocation}. 
Recent repository understanding approaches extend this idea to large-scale codebases by constructing feature-oriented representations and traceability links. 
For example, RepoSummary~\cite{RepoSummary} extracts repository summaries organized around functional features and associates feature descriptions with relevant implementation components.
These studies demonstrate that a feature is rarely implemented by a single function; instead, realizing a functionality often requires locating a set of interacting and distributed code entities.

More recently, features have become natural task units in LLM-based repository-level software engineering. 
Benchmarks such as FEA-Bench~\cite{li2025-feabench}, FeatBench~\cite{FeatBench}, NoCode-bench~\cite{NoCode-bench}, and FeatureBench~\cite{FeatureBench} evaluate whether LLM-based systems can implement concrete feature requests in existing repositories. 
Beyond evaluation, SWE-Dev~\cite{du2026SWE-DEV} and EvoDev~\cite{EvoDev} investigate feature-driven autonomous development and dependency modeling, while FeatX~\cite{FeatX} supports repository evolution through feature-level editing interactions.
Together, these studies establish features as meaningful units for repository-level development. 
However, they mainly treat features as task specifications, evaluation targets, or editing units, rather than as retrieval signals for identifying reusable code context.

FeatLens builds on this perspective by treating feature-level intent as an operational retrieval signal. 
It maps feature descriptions to supporting repository entities and uses them as dependency seeds for constructing task-specific code context, thereby connecting functional requirements with the code dependencies needed for implementation.

%% file: chapter/09-Conclusion.tex
\section{Conclusion}

This paper introduced FeatLens, a feature-guided dependency retrieval approach for repository-level code generation.
FeatLens maps feature-level intent to function-level dependency context through Feature-Oriented Repository Indexing, Feature Retrieval and Dependency Seeding, and Semantic-Structural Dependency Context Refinement. Retrieval experiments on DevEval and EvoCodeBench show that FeatLens achieves the highest dependency recall among the evaluated methods at each fixed context budget, without online LLM retrieval calls. Downstream generation experiments on DevEval yield the highest non-oracle Pass@1 point estimates and the shortest generated completions under both DeepSeek-V3.2 and GPT-5-mini.
Future work will strengthen dependency context refinement for runtime binding, implicit framework conventions, and dependencies that are weakly aligned with feature clusters, and broaden evaluation with more diverse repository-level code generation benchmarks.